\documentclass[12pt]{iopart}
\usepackage{graphicx}
\usepackage{here}

\begin{document}

\title[Resonating Valence
Bond physics in spin-orbital models]{The emergence of Resonating Valence
Bond physics in spin-orbital models}

\author{F Mila$^1$, F Vernay$^2$, A Ralko$^1$, F Becca$^3$, P Fazekas$^4$, K Penc$^4$}
\address{$^1$ Institute of Theoretical Physics,
Ecole Polytechnique F\'ed\'erale de Lausanne,
CH-1015 Lausanne, Switzerland} 
\address{$^2$ Department of Physics, University of Waterloo, Waterloo, ON, N2L3G1, Canada}
\address{$^3$ INFM-Democritos, National Simulation Center and International School for
Advanced
Studies (SISSA), I-34014 Trieste, Italy}
\address{$^4$ Research Institute for Solid State
Physics and Optics, H-1525 Budapest, P.O.B. 49, Hungary}
\ead{frederic.mila@epfl.ch}
\begin{abstract}
We discuss how orbital degeneracy, which is usually
removed by a cooperative Jahn-Teller distortion, could
under appropriate circumstances lead rather to a Resonating Valence Bond
spin-orbital liquid. The key points are: i) The tendency to form
spin-orbital dimers, a tendency already identified in several
cases; ii) The mapping onto Quantum Dimer Models,
which have been shown to possess Resonating Valence Bond 
phases on the triangular lattice. How this program can be 
implemented is explained in some details
starting from a microscopic model of LiNiO$_2$.
\end{abstract}

\pacs{75.10.Jm, 05.50.+q, 05.30.-d}

\section{Introduction}
It is very useful to classify quantum magnets
according to the symmetry (if any) that is
broken in the ground state. When the SU(2) symmetry
is broken, the system usually sustains some kind of long-range
magnetic order, although some more exotic examples
involving quadrupolar order have been recently discussed\cite{arikawa,laeuchli}.
When the SU(2) symmetry is not broken, a translation symmetry
may or may not be broken depending on the lattice topology.
The standard case in which no lattice symmetry is broken is
that of systems in which it is possible to build  a singlet inside
the unit cell, as for instance in a ladder\cite{dagotto}. If however this is not possible,
as in all systems with half-integer spins and an odd number
of sites per unit cell, the simplest way to keep SU(2) symmetry
is to break the translation symmetry so that the new unit cell
contains an even number of sites per unit cell. The typical example
is the S=1/2 $J_1-J_2$ chain, which has been explicitly shown 
quite some time ago by Majumdar and Ghosh\cite{majumdar} to have two-fold 
degenerate dimerized ground state when $J_2=J_1/2$.

Another possibility to keep SU(2) symmetry without breaking 
any lattice symmetry has been put forward by Anderson\cite{anderson} in 1973.
Concentrating on S=1/2 magnets, he suggested that under 
appropriate circumstances the ground state might be a linear 
combination of valence bond (VB) states, i.e. states in which sites
are paired to form singlets. Clearly each individual state breaks
at least some of the translational symmetries of the underlying
lattice, but the translational symmetry is restored by the
superposition of such valence bond states. Such a wave
function is known as a Resonating Valence Bond (RVB) state.

To identify such a ground state in realistic models of quantum
antiferromagnets turned out to be much more difficult than
anticipated. The prediction of Fazekas and Anderson\cite{fazekas} that this
might be the case for the S=1/2 Heisenberg model on the triangular
lattice, based on estimates of the energy of ordered and valence
bond states including perturbation corrections, is not supported
by recent numerical investigations, which all point to 3-sublattice
antiferromagnetic order\cite{bernu}. 

The most serious candidate still around is the S=1/2 Heisenberg
antiferromagnet on the kagome lattice. No evidence of magnetic
long-range order could be found so far, and the proliferation of 
low-lying singlets observed in exact diagonalizations of finite
clusters\cite{lecheminant} can be fairly well described in the short-range RVB 
subspace of valence bond states involving only nearest neighbours\cite{mambrini}.
However, several treatments based on some effective Hamiltonian
have reached the conclusion that the ground state support some 
kind of valence bond order\cite{syromyatnikov,nikolic,auerbach}, and hence breaks 
translational symmetry.
Unfortunately, the resulting unit cell is so large, and the singlet-singlet
gap accordingly so small, that this prediction cannot be cross-checked
by the only unbiased numerical approach available so far, namely
exact diagonalization, and the issue is likely to remain open for quite some
time.

In fact, it has only been possible so far to  unambiguously identify
an RVB ground state in a very minimal description of fluctuations
in the RVB subspace that goes under the name of Quantum Dimer
Model (QDM)\cite{rokhsar}. In this approach, valence bond configurations are assumed
to build an orthogonal basis of the Hilbert space, and the effective
Hamilonian contains kinetic terms that shift dimers around loops
and potential terms that favour or penalize specific local configurations
of dimers. For the triangular lattice, the simplest model is defined
by the Hamiltonian:

\begin{figure}[H]
\newcommand{\lb}[1]{\raisebox{-0.8ex}[0.8ex]{#1}}
\begin{center}$H = v \sum \big(\, |$
\lb{\resizebox{0.035\textwidth}{!}{
\includegraphics[height=5cm]{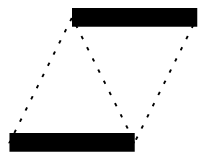}}} $\rangle\,\langle$
\lb{\resizebox{0.035\textwidth}{!}{
\includegraphics[height=5cm]{fig1.eps}}} $| + |$
\lb{\resizebox{0.035\textwidth}{!}{
\includegraphics[height=5cm]{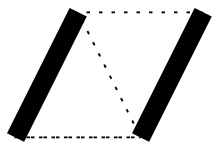}}} $\rangle\,\langle$
\lb{\resizebox{0.035\textwidth}{!}{
\includegraphics[height=5cm]{fig2.eps}}} $|\, \big)$ 
$ \ - t  \sum \big(\, |$
\lb{\resizebox{0.035\textwidth}{!}{
\includegraphics[height=5cm]{fig1.eps}}} $\rangle\,\langle$
\lb{\resizebox{0.035\textwidth}{!}{
\includegraphics[height=5cm]{fig2.eps}}} $| + |$
\lb{\resizebox{0.035\textwidth}{!}{
\includegraphics[height=5cm]{fig2.eps}}} $\rangle\,\langle$
\lb{\resizebox{0.035\textwidth}{!}{
\includegraphics[height=5cm]{fig1.eps}}} $|\, \big)$
\end{center}
\end{figure}
\noindent where the sum runs over all plaquettes including
the three possible orientations.  
The kinetic term controlled by the amplitude $t$ changes the dimer covering
of every flippable plaquette, i.e., of every plaquette containing two dimers
facing each other, while the potential term controlled by the interaction $v$
describes a repulsion ($v>0$) or an attraction ($v<0$) between dimers
facing each other. Since a positive $v$ favors configurations without flippable plaquettes
while a negative $v$ favors configurations with the largest possible number of 
flippable plaquettes, one might expect a phase transition between two phases as a function of $v/t$. 
The actual
situation is far richer though. As shown by Moessner and Sondhi\cite{moessner}, who calculated the 
temperature 
dependence of the structure factor, there are four different
phases: {\bf i)} A staggered phase for $v/t>1$, 
in which the ground-state manifold consists of all non-flippable configurations;
{\bf ii)} A columnar ordered phase for $v/t$ sufficiently negative; {\bf iii)} An ordered phase 
adjacent to it  which probably consists of resonating
plaquettes which make a $12$-site unit-cell pattern\cite{ralko2}; 
{\bf iv)} A liquid phase with a featureless and temperature independent 
structure factor. This last phase has been interpreted as a short-range 
(RVB) phase in which all correlations decay exponentially at zero temperature,
an interpretation confirmed by recent Green's function Quantum Monte
Carlo simulations\cite{ralko1}, which have established the presence of topological 
degeneracy in this parameter range, a clear characteristic of the RVB phase.

\begin{figure}
\begin{center}
 \vspace{0.3cm}
 \includegraphics[width=0.3\textwidth]{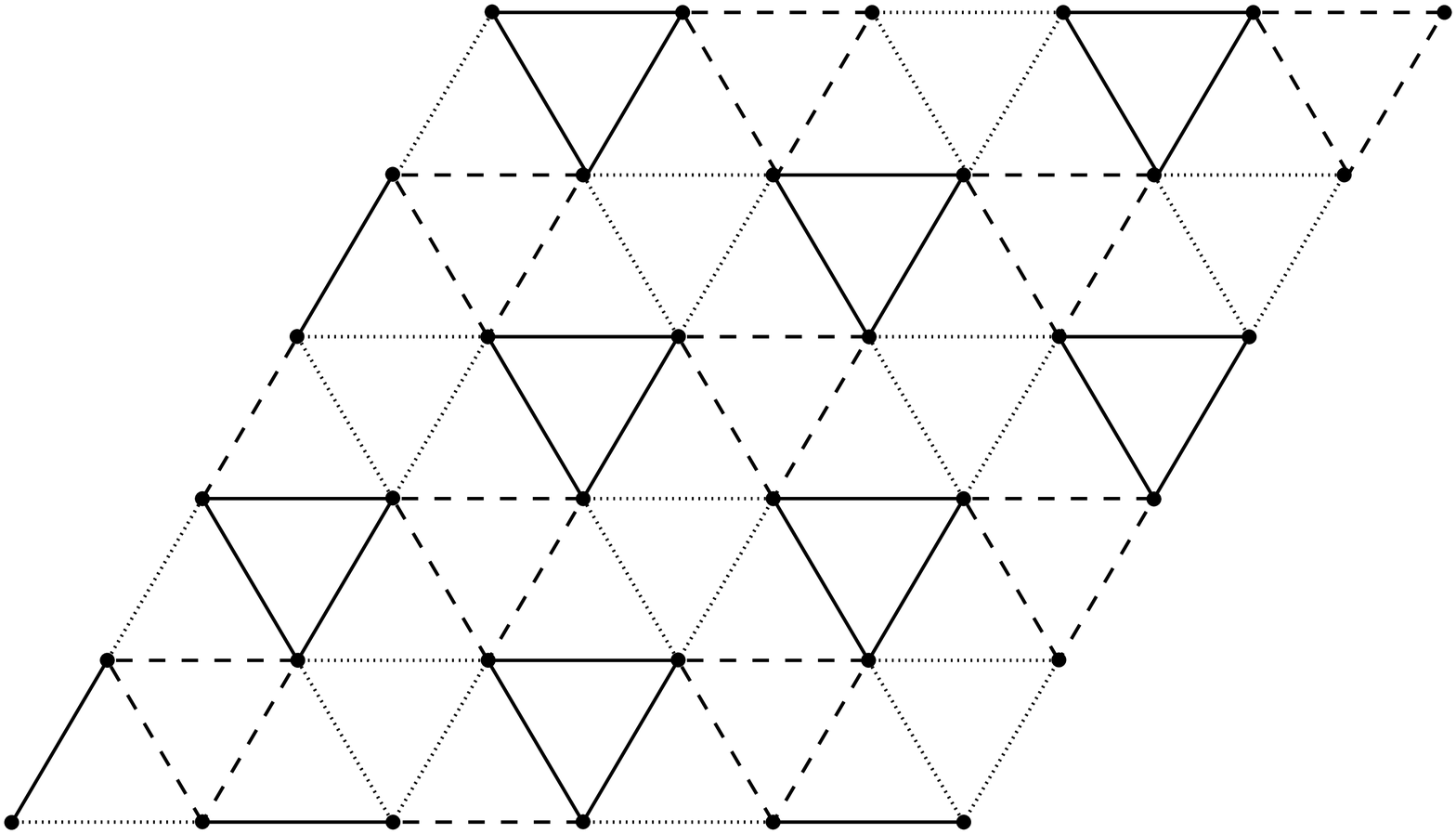}
 \caption{\label{fg:triankind}
    Triangular lattice on which the spin-chirality Hamiltonian is defined.
    The unitary vector for the bond is indicated by solid lines
    (${\vec e}_\mu = {\vec e}_1$), dashed lines (${\vec e}_\mu = {\vec
      e}_2$), and dotted lines (${\vec e}_\mu = {\vec e}_3$).}
\end{center}
\end{figure}

This result defines a new line of research in the field: Indeed, rather than investigating
directly the properties of the Heisenberg model on a given lattice, one can
try to identify models for which a VB subspace is a reasonable variational
subspace, derive an effective QDM, and study it
along the same lines as the minimal model on the triangular lattice.
In that respect, a natural candidate is the trimerized spin-1/2 Heisenberg
model on the kagome lattice. An effective model in terms of the total spin
$\vec \sigma$ and a chirality pseudo-spin $\vec \tau$ per strong triangle has 
been derived\cite{subra,mila}. It is defined on the triangular lattice built by strong triangles
and can be written\cite{ferrero}:
\begin{equation}
  {\cal H}_0^{{\rm eff}} = \frac{J^\prime}{9}
  \sum_{\langle i,j \rangle}
  {\vec\sigma}_i \cdot {\vec\sigma}_j
  (1 - 4 {\vec e}_{ij} \cdot {\vec \tau}_i)
  (1 - 4 {\vec e}_{ij} \cdot {\vec \tau}_j),
\end{equation}
where $J'$ is the weak coupling between the strong triangles of the
trimerized lattice, and where the vectors ${\vec e}_{ij}$ have
to be chosen among 
${\vec e}_1=(1,0)$, ${\vec e}_2=(-\frac{1}{2},-\frac{\sqrt{3}}{2})$,
${\vec e}_3=(-\frac{1}{2},\frac{\sqrt{3}}{2})$
according to the pattern of Fig.~\ref{fg:triankind}.
A mean-field decoupling of spin and chirality has identified nearest-neighbour
valence bond states as the lowest solutions\cite{mila}, and, following Rokhsar and 
Kivelson\cite{rokhsar}, a QDM
has been derived\cite{zhitomirsky}. Unfortunately, the properties of this model could not 
be studied so far. First of all, they involve kinetic terms on longer loops than
the above-mentioned minimal model, but more importantly, it is impossible
to formulate it in such a way that all off-diagonal matrix elements are negative,
so that the Quantum Monte Carlo methods that were successful for the minimal
model cannot be used.

The effective spin-chirality model has another remarkable feature though:
It is formally very similar to the spin-orbital models that are used to describe
Mott insulators with orbital degeneracy. Indeed, as discussed in great details
by Kugel and Khomskii\cite{kugel}, when the local symmetry is such that different
orbitals can be occupied in the open shell of the magnetic ions of a Mott
insulator, this extra degree of freedom can be described by a pseudo spin,
and the resulting model is roughly speaking of the same form. Given the very different
precise forms this model can take for specific systems, a general discussion
cannot be attempted here. Rather, we concentrate in the next section on
a specific example of Mott insulator with orbital degeneracy for which we believe that RVB physics might
be realized. More general comments will be given in the last two sections 
of the manuscript.

\section{The spin-orbital model of LiNiO$_2$}
\begin{figure}[ht]
\begin{center}
\includegraphics*[width=12truecm,angle=0]{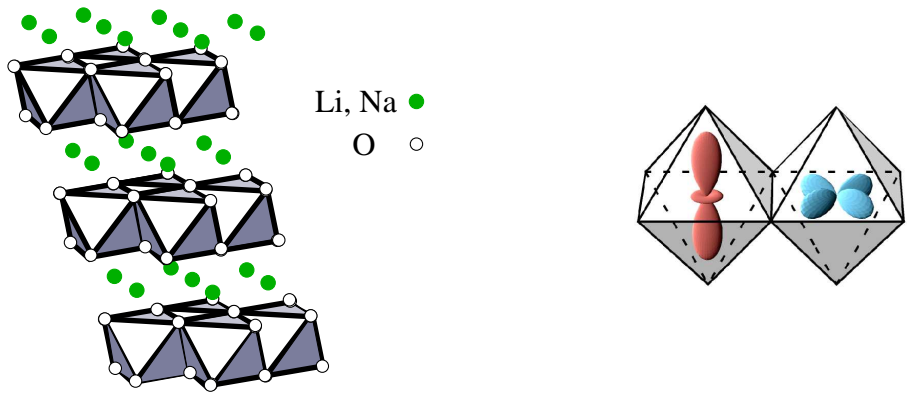}
\caption[ANiO2]{\label{ANiO2} 
Left: ANiO$_2$ structure. Ni ions are located in the middle of the O octahedra.
Right: Local structure and degenerate orbitals: $d_{3z^2-r^2}$ (left octahedron) and $d_{x^2-y^2}$ 
(right octahedron)}
\end{center}
\end{figure}

The Mott insulator LiNiO$_2$ and its cousin NaNiO$_2$ 
are isostructural and isoelectronic. The crystal structure can be envisaged 
as a sequence of slabs of edge sharing 
octahedra of oxygen O$^{2-}$ ions. Metal ions sit at the centers of
octahedra. There are two kinds of slabs: in A slabs, at every center of 
octahedra there is a Ni$^{3+}$, whereas in the B slabs, one finds either 
Li$^+$ or Na$^+$ ions. A and B slabs alternate (see Fig.~\ref{ANiO2}). 
The Ni ions form well-separated triangular planes.  
The Ni$^{3+}$ ions are in the $S=1/2$ low-spin state, which allows for 
twofold orbital degeneracy between the $d_{3z^2-r^2}$ and $d_{x^2-y^2}$
orbitals (see Fig.~\ref{ANiO2}).

Surprisingly enough, the two systems have very different properties:
NaNiO$_2$ undergoes a high temperature Jahn-Teller distortion,
followed at low temperature by the antiferromagnetic ordering of
ferromagnetic triangular planes, which makes it a standard example
of orbital ordering followed by magnetic ordering\cite{nanio2exp}. 
By contrast,
no ordering could be detected in LiNiO$_2$, and some kind of 
freezing seems to take place below 8 K\cite{linio2exp}. The best samples of LiNiO$_2$
are non stoichiometric however, some  Li sites being occupied by Ni atoms, and this has been 
invoked to explain the difference. The trend upon approaching 
stoichiometry is not clear though, and this striking difference between
the two systems, in particular the lack of any sign of a cooperative Jahn-Teller transition in 
LiNiO$_2$, suggests to look for alternative explanations. In the 
following, starting from a realistic microscopic
description of the system, we study the possibility to stabilize an RVB 
spin-orbital liquid  in the absence of any disorder.

\subsection{Microscopic Model}

A fairly general description of this system is given by
a Kugel-Khomskii Hamiltonian defined in terms of Wannier functions centered 
on the Ni sites by two hopping integrals $t_h$ 
and $t^\prime_h$, the on-site Coulomb repulsion $U$ and the Hund's coupling
$J$ which, on a given bond, takes the form\cite{vernay1}
\begin{eqnarray}
\mathcal{H}_{ij} =
  -\frac{2}{U\!+\!J} \left[
    2 t_h t^\prime_h {\bf T}_i {\bf T}_j
  - 4 t_h t^\prime_h T^y_i T^y_j 
    + (t_h-t^\prime_h)^2 ({\bf n}_{ij}^z{\bf T}_i) ({\bf n}_{ij}^z{\bf T}_j)\right.\nonumber\\ 
+ \frac{1}{2}  \left. (t_h^2-{t^\prime_h}^2) \left( {\bf n}_{ij}^z{\bf T}_i
+ {\bf n}_{ij}^z{\bf T}_j \right) +  \frac{1}{4}(t_h^2+ {t^\prime_h}^2)  
    \right] \mathcal{P}_{ij}^{S=0} 
\nonumber\\
  -\frac{2}{U -J} \left[
    4 t_h t^\prime_h T^y_i T^y_j 
  + \frac{1}{2}(t_h^2+ {t^\prime_h}^2)  
  + \frac{1}{2} (t_h^2-{t^\prime_h}^2) \left(
  {\bf n}_{ij}^z{\bf T}_i + {\bf n}_{ij}^z {\bf T}_j 
\right)
\right] \mathcal{P}_{ij}^{S=0} 
\nonumber\\
  -\frac{2}{U\!-\!3J} \left[  - 2 t_h t^\prime_h {\bf T}_i {\bf T}_j
   - 
   (t_h-t^\prime_h)^2 ({\bf n}_{ij}^z{\bf T}_i) ({\bf n}_{ij}^z{\bf T}_j)+
\frac{1}{4}(t_h^2+{t^\prime_h}^2)
    \right] \mathcal{P}_{ij}^{S=1} 
\label{eq:effham3}
\end{eqnarray}
with the usual definitions for the projectors on the singlet and triplet states
of a pair of spins:
\begin{equation}
\mathcal{P}_{ij}^{S=0} = \frac{1}{4} - {\bf S}_i{\bf S}_j 
\quad\mbox{and}\quad
\mathcal{P}_{ij}^{S=1} = {\bf S}_i{\bf S}_j+\frac{3}{4},
\end{equation}
The vectors ${\bf n}_{ij}^z$ depend on the orientation of the bonds and are given by
${\bf n}_{12}^z=(0,0,1)$,
${\bf n}_{13}^z=(\frac{\sqrt{3}}{2},0,-\frac{1}{2})$ and
${\bf n}_{23}^z=(-\frac{\sqrt{3}}{2},0,-\frac{1}{2})$ for the 3 orientations respectively.
The operators ${\bf T}_i$ are pseudo-spin operators acting on the orbitals.
For the local geometry shown in the right panel of Fig.~\ref{ANiO2}, $t_h$ and $t'_h$ correspond to the hopping between
pairs of $d_{3z^2-r^2}$ and $d_{x^2-y^2}$ respectively, the hopping between
the $d_{3z^2-r^2}$ on one site and the $d_{x^2-y^2}$ on the other being zero by symmetry.
Of course, all bonds are equivalent, but once a basis, i.e. a pair of local orbitals, 
has been chosen, the Hamiltonian takes a different form on bonds with different
orientations. 
Note that other forms of the microscopic Hamiltonian including explicitly O orbitals
have been used\cite{dare,mostovoy,reitsma} with somewhat different conclusions.

\subsection{Mean-field phase diagram}

\begin{figure}[ht]
\begin{center}
\includegraphics*[width=8truecm,angle=0]{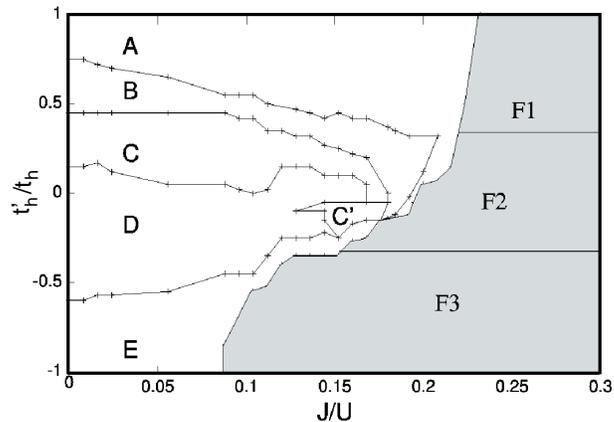}
\caption[var-dia]{\label{var-dia}{Mean-field phase diagram on a 16-site cluster
as a function of hopping 
integral versus Hund's coupling. The grey phase is the ferromagnetic phase,
with the classical phase boundaries between different types of orbital ordering.}}
\end{center}
\end{figure}

Inspired by the results obtained on the trimerized kagome model\cite{mila}, 
spin and orbital degrees of 
freedom can be decoupled in a  mean-field way\cite{vernay1}.
This leads to a phase diagram in which phases can be distinguished
by the mean value of the orbital and/or spin part of the Hamiltonian
on each bond. The resulting phase diagram is remarkably rich (see
Fig.~\ref{var-dia}). Orbital ordering in the ferromagnetic phase has also been discussed
in Ref.\cite{mostovoy}.

\begin{figure}[ht]
\begin{center}
\includegraphics*[width=12truecm,angle=0]{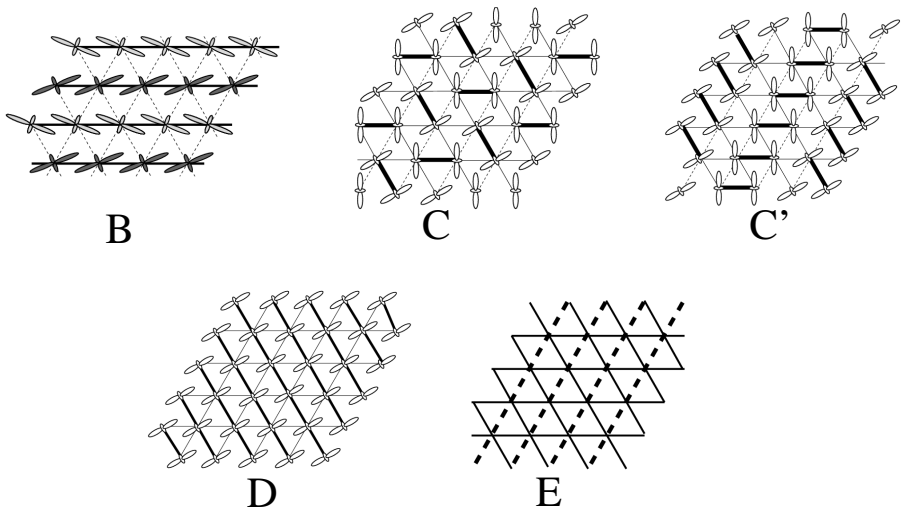}
\caption[var-dia]{\label{phases}{Spin and orbital structure in the singlet
phases of the mean-field phase diagram. Solid line indicates AF, dashed line
FM spin correlations.}}
\end{center}
\end{figure}

While the planes of NaNiO$_2$ are known by now to
be ferromagnetic, suggesting that NaNiO$_2$ is in one of the ferromagnetic phases,
 LiNiO$_2$ is expected to be in one of the antiferromagnetic phases.
The orbital and spin structure of the antiferromagnetic phases is depicted
in Fig.~\ref{phases}, except phase A, which consists of SU(4) plaquettes\cite{penc1}
and cannot be described along these lines.
Since niether orbital nor magnetic ordering could be detected in LiNiO$_2$, and since
phases B and  D have a simple orbital ordering pattern while phase E is 
likely to be antiferromagnetically ordered, let us concentrate
on phases C and C'.
Both phases are characterized by strong dimer singlets 
forming different regular dimer coverings of the triangular 
lattice. On each dimer the orbitals are parallel, and
they correspond to $d_{3z^2-r^2}$, $d_{3x^2-r^2}$ or 
$d_{3y^2-r^2}$ depending on the orientation of the bond.
Note that all these orbitals are Jahn-Teller active,
leading in all cases to two long and four 
short Ni-O bonds.
One might be tempted to conclude that these phases
correspond to two types of valence bond solids with 
the patterns depicted in Fig.~\ref{phases}. The mean-field
approach has a very remarkable property however: In 
addition to the mean-field solutions with lowest energy shown
in Fig.~\ref{phases}, there are several other
mean-field solutions of the self-consistent equations
with energies very close to the lowest energy corresponding
to other dimer coverings of the triangular lattice\cite{BaVS3}.
In such circumstances, going beyond mean-field is likely
to couple these solutions, and the relevant model would
then be a QDM describing resonances between
these states, a point of view
favoured by exact diagonalizations of finite clusters.
So it is more appropriate to think of these
phases as a region of parameters where all dimer coverings
are relevant states for low-energy physics.

\subsection{Effective Quantum Dimer Model}

Starting from all dimer configurations mentioned in the previous section,
a QDM has been derived\cite{vernay2}. It involves a competition
between kinetic processes and dimer-dimer repulsion. A miminal
version of the model is defined by:
\begin{equation}\label{hamilt}
\begin{array}{rcl}
{ H}= \ &-&t \sum 
\left(
|\unitlength=1mm
\begin{picture}(6.2,5)
\linethickness{2mm}
\put(0.9,-.7){\line(1,2){1.8}}
\put(3.8,-.7){\line(1,2){1.8}}
\end{picture}
\rangle
\langle
\unitlength=1mm
\begin{picture}(6.5,5)
\linethickness{0.3mm}
\put(3.2,2.6){\line(1,0){3.2}}
\put(0.9,-.7){\line(1,0){3.2}}
\end{picture}
|
+h.c.\right)
- \ t^\prime \sum 
\left(
\left|\unitlength=1mm
\begin{picture}(7,6)
\linethickness{2mm}
\put(0.8,-1.7){\line(1,2){1.8}}
\put(6.4,1.6){\line(-1,2){1.8}}
\linethickness{0.2mm}
\put(3.8,-1.7){\line(1,0){3.6}}
\end{picture}
\right\rangle
\left\langle
\unitlength=1mm
\begin{picture}(7,6)
\linethickness{2mm}
\put(1.8,1.6){\line(1,2){1.8}}
\put(7,-1.7){\line(-1,2){1.8}}
\linethickness{0.2mm}
\put(-0.6,-1.7){\line(1,0){3.6}}
\end{picture}
\right|
+h.c.\right)\nonumber \\

&+& \ V \sum \left(
|\unitlength=1mm
\begin{picture}(6.2,5)
\linethickness{2mm}
\put(0.9,-.7){\line(1,2){1.8}}
\put(3.8,-.7){\line(1,2){1.8}}
\end{picture}
\rangle
\langle
\unitlength=1mm
\begin{picture}(6.2,5)
\linethickness{2mm}
\put(0.9,-.7){\line(1,2){1.8}}
\put(3.8,-.7){\line(1,2){1.8}}
\end{picture}|+
|
\unitlength=1mm
\begin{picture}(6.5,5)
\linethickness{0.3mm}
\put(3.2,2.6){\line(1,0){3.2}}
\put(0.9,-.7){\line(1,0){3.2}}
\end{picture}\rangle
\langle
\begin{picture}(6.5,5)
\linethickness{0.3mm}
\put(3.2,2.6){\line(1,0){3.2}}
\put(0.9,-.7){\line(1,0){3.2}}
\end{picture}
|
\right),
\end{array}
\end{equation}
where the sums run over the 4-site and 6-site loops with all possible 
orientations. Although the repulsion is a higher order process, hence quite small, 
and although the ratio $t^\prime/t$ is in principle fixed by the 
perturbative expansion, these parameters are treated as free
to make contact with the Rokhsar-Kivelson model
on the triangular lattice. The main difference with the effective model derived 
for the trimerized 
kagome antiferromagnet is that the off-diagonal elements are now all {\it negative}. This 
is in practice extremely important since it allows one to use Quantum Monte
Carlo simulations.

\begin{figure}
\begin{center}
\includegraphics[width=0.40\textwidth]{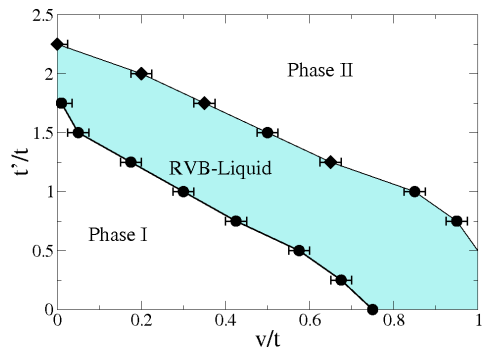}
\caption{\label{fig:phasediag} 
(Color online) Phase diagram in the $t^\prime{-}V$ plane. A wide disordered region
extends all the way from the standard QDM ($t^\prime/t=0$ axis) to the purely
kinetic QDM ($V/t=0$ axis). The description of the symbols is given in the text.}
\end{center}
\end{figure}

The phase diagram of the model has been derived using exact diagonalizations
of finite clusters and Green's function Quantum Monte Carlo\cite{vernay2}. As shown before,
the most convenient way to identify an RVB phase is to look for topological 
degeneracy since it is at least partially lifted in all other phases. The resulting
phase diagram is shown in Fig.~\ref{fig:phasediag}. It contains a large RVB liquid phase
which connects the relevant parameter range for LiNiO$_2$ ($t'/t \simeq 2$ and $V$ small)
to the RVB liquid phase of the minimal model ($t'=0$). Translated into spin-orbital
language, this RVB phase corresponds to a spin-orbital liquid wiht
no symmetry breaking and no phase transition, in agreement with 
the phenomenology of LiNiO$_2$.

\section{Discussion}

Let us now comment on how generic the mechanism proposed in the context
of the spin-orbital model of LiNiO$_2$ might be. The main ingredients to get 
an RVB spin-orbital liquid are: 1) The spontaneous formation of dimers; 
2) The degeneracy or
quasi-degeneracy of the energies of the wave-functions constructed out of
different dimer coverings; 3) The presence of an RVB phase in the relevant
QDM. Let us comment on these points separately.

The tendency of spin-orbital models to spontaneously form
dimers is well documented. The possibility to stabilize dimerized
 ground states due to orbital
degeneracy has first been put forward by Feiner {\it et al} in the context of 
a realistic 3D model\cite{feiner}. Shortly after, the presence of a dimerized
ground state has been explicitly proven  for a simple minimal 1D model by Kolezhuk 
and Mikeska\cite{kolezhuk}, a result generalized shortly after by Pati {\it et al}\cite{pati} 
in the context of a model defined by the Hamiltonian:

\begin{equation}
H=\sum_{\langle i,j \rangle} \left[J_1 \vec S_i.\vec S_j +J_2 \vec T_i.\vec T_j + K(\vec S_i.\vec S_j )(\vec T_i.\vec T_j )\right]
\end{equation}
with $K>0$. When $J_1/K=J_2/K=3/4$, it can be rewritten as

\begin{equation}
H=K\sum_{\langle i,j \rangle}  (\vec S_i.\vec S_j +3/4)(\vec T_i.\vec T_j + 3/4)
\end{equation}
Each term is obviously positive, and since $\vec S_i.\vec S_j +3/4$ 
(resp. $\vec T_i.\vec T_j + 3/4$) is the projector on the spin (resp. orbital) triplet,
the two wave-functions with alternating spin and orbital singlets are zero energy 
eigenstates, hence ground-states\cite{kolezhuk}. Pati {\it et al} have shown that this dimerized
phase extends to a very large portion of the phase diagram around this point. 
From that point of view, the identification of dimer phases in the context of spin-orbital
models of BaVS$_3$\cite{BaVS3} and of LiNiO$_2$\cite{vernay1} is
not unexpected, and the tendency to dimerize can be considered to be a generic 
trend of spin-orbital models.

What seems to be more specific to these spin-orbital models of LiNiO$_2$
and BaVS$_3$ is the quasi-degeneracy of all nearest-neighbour dimer
coverings. But in fact, this can be traced back to a rather generic feature of 
spin-orbital models, namely the remarkable symmetry properties of the orbital part
of the Hamiltonian. As can be clearly seen in the spin-orbital model of LiNiO$_2$, the orbital
part does not have the same form in the three directions of the triangular lattice,
a property encoded in the $\bf n_{ij}^z$ vectors. So the Hamiltonian is only 
invariant if one simultaneously performs the same rotation in real space
and in pseudo-spin space. For purely orbital models, this is known to have
remarkable consequences\cite{nussinov,ma,doucot,dorier}. For spin-orbital
models, this implies that dimers with different orientations involve different
orbital wave-functions, as can be clearly seen in phases C and C'. What
controls the energy of a given dimer configuration is then the residual dimer-dimer
interaction.  It turns out  that simple patterns having
all dimers parallel to each other are not naturally favoured if the anisotropy of the orbital
part is strong because it is impossible to gain energy in the other directions.
On a lattice such as the triangular lattice with a large connectivity, it is then
much more 
favourable to adopt configurations where dimers are not parallel to each other.
The energy difference between such configurations however is not really 
significant, and it is better to look at such states as a variational basis.

Finally, too little is known at that stage about RVB phases in QDM to draw
general conclusions, but it seems plausible that the presence of an RVB phase
between two valence-bond phases is the generic alternative to a first order
transition.

To summarize, the tendency toward dimerization is a rather general feature
of spin-orbital models.
When confronted to a lattice such as the triangular lattice, for which 
QDM's are known by now to possess RVB phases, there
is a real chance for quantum fluctuations to stabilize an RVB ground
state.

\section{Conclusion}

We have shown that orbital degeneracy can, under special but
neither unrealistic nor fine-tuned conditions, lead to a spin-orbital
RVB ground state. Clearly, this is not the most common situation. 
Indeed, orbital degeneracy usually leads to a cooperative Jahn-Teller
transition, resulting into an effective spin Hamiltonian with a symmetry 
different from that of the original lattice\cite{kugel}. This is not either the
only route to spin-orbital liquid behaviour\cite{loidl}.
However, the tendency of spin-orbital models to spontaneously dimerize
is strong enough to make this a promising route towards RVB physics.
Whether LiNiO$_2$ is the first example remains to be seen. To make progress
on this issue will require not only further theoretical work
to better understand the relevant microscopic model and its possible
connection to a QDM, but also and maybe more importantly
further experimental investigations to unambiguously identify
the orbital and magnetic properties of stoichiometric samples.

We acknowledge useful discussions with M. Ferrero and D. Ivanov.
This work was supported by the Swiss National Fund and by MaNEP.

\end{document}